\documentclass[aps,prd,preprint,preprintnumbers,nofootinbib,showpacs,floatfix,amsmath,amssymb]{revtex4}

\usepackage{graphicx}
\usepackage{subfig}
\usepackage{wasysym} 

\newcommand{\DOEnum}{DE-SC0008347}
\newcommand{\support}{This work was supported by the U.S. Department of Energy under award No.~\DOEnum.}

\usepackage{mathrsfs}

\newcommand{\mg}{\textsc{MadGraph}}
\newcommand{\mgCite}{{\mg}5 v{\mgFiveVersion} \cite{MadGraph:5}}

\newcommand{\pythia}{\textsc{Pythia}}
\newcommand{\pythiaCite}{{\pythia}~{\pythiaEightVersion} \cite{Pythia:8.1, Pythia:physics}}

\newcommand{\delphes}{\textsc{Delphes}}
\newcommand{\delphesCite}{{\delphes}~3.2 \cite{Delphes:3}}

\newcommand{\fastjet}{\texttt{FastJet}}
\newcommand{\fastjetCite}{{\fastjet}~{\fastjetVersion} \cite{FastJet:user-manual}}

\newcommand{\ds}{^{}} 

\newcommand{\pT}{p_{T}\ds}

\newcommand{\HT}{H_{T}\ds}
\newcommand{\antiKt}{anti-$k_{T}\ds$}

\newcommand{\etaJet}[1][]{\eta_{\mathrm{jet}}^{#1}}

\newcommand{\TeV}{\mathrm{TeV}}

\newcommand{\SUtwo}[1][]{\textrm{SU(2)}_{\textrm{#1}}}

\newcommand{\Kshort}{K_{S}^{0}}
\newcommand{\Klong}{K_{L}^{0}}

\newcommand{\lagrange}{\mathscr{L}}
\newcommand{\invfb}{\mathrm{fb}^{-1}}
\newcommand{\lambdaQCD}{\Lambda_{\mathrm{QCD}}}

\newcommand{\bigO}[1]{\mathcal{O}(#1)}
\newcommand{\abs}[1]{\left|#1\right|}
\newcommand{\pair}[1]{#1\bar{#1}} 
\newcommand{\about}[1]{{\sim}#1} 

\newcommand{\powerUnit}[2]{10^{#1}\,\mathrm{#2}}

\newcommand{\new}         {$\mu_{x}\ds$} 
\newcommand{\runII}       {Run II} 
\newcommand{\runIIlum}    {100\,\invfb} 
\newcommand{\minJetPt}    {300} 
\newcommand{\pileupMu}    {40} 
\newcommand{\deltaEtaMax} {1.5} 

\newcommand{\minMuPt}       {10\textrm{ GeV}} 
\newcommand{\minTowerf}     {0.05} 
\newcommand{\jetR}          {0.4} 
\newcommand{\coreR}         {0.04} 
\newcommand{\coreMass}      {2\textrm{ GeV}} 
\newcommand{\BMass}         {5.3\textrm{ GeV}} 
\newcommand{\maxSubjetMass} {12\textrm{ GeV}} 
\newcommand{\cutX}          {3} 
\newcommand{\cutXrho}       {90\%} 
\newcommand{\minSubjetF}    {0.5} 

\newcommand{\mgFiveVersion}       {2.2.3} 
\newcommand{\sqrtS}               {\sqrt{S}=13\textrm{ TeV}}  
\newcommand{\pdfSetCite}          {CT14llo PDFs \cite{CTEQ:CT14}} 
\newcommand{\pythiaEightVersion}  {8.210} 
\newcommand{\fastjetVersion}      {3.1.2} 

\newcommand{\pTrm}[1]   {p_{T, \mathrm{#1}}\ds}
\newcommand{\pTmu}      {p_{T,\mu}\ds}
\newcommand{\pTrel}     {p_{T}^{\mathrm{rel}}}
\newcommand{\pTfrac}[2] {f_{\mathrm{#1}}^{\mathrm{#2}}}

\newcommand{\pSubjet}[1][] {p_{\mathrm{subjet}}^{#1}}
\newcommand{\pCore}[1][]   {p_{\mathrm{core}}^{#1}}
\newcommand{\pMu}[1][]     {p_{\mu}^{#1}}
\newcommand{\pNu}[1][]     {p_{\nu_{\mu}}^{#1}}

\newcommand{\gammaB}         {\gamma_{B}\ds}
\newcommand{\gammaMuCM}      {\gamma_{\mu, \mathrm{cm}}\ds}
\newcommand{\gammaCore}[1][] {\gamma_{\mathrm{core}}^{#1}}
\newcommand{\gammaSubjet}    {\gamma_{\mathrm{subjet}}\ds}
\newcommand{\mB}             {m_{B}\ds}
\newcommand{\mCore}[1][]     {m_{\mathrm{core}}^{#1}}
\newcommand{\mSubjet}        {m_{\mathrm{subjet}}\ds}
\newcommand{\betaMuCM}       {\beta_{\mu,\mathrm{cm}}}
\newcommand{\thetaCM}        {\theta_{\mathrm{cm}}\ds}
\newcommand{\thetaLab}       {\theta_{\mathrm{lab}}\ds}
\newcommand{\fSubjet}        {\pTfrac{subjet}{}}

\newcommand{\xCut}[1][]   {x_{\mathrm{max}}^{#1}} 
\newcommand{\xiCut}[1][]  {\xi_{\mathrm{max}}^{#1}} 
\newcommand{\fCut}        {\pTfrac{subjet}{min}}
\newcommand{\Rcore}       {R_{\mathrm{core}}\ds}
\newcommand{\mSubjetMax}  {m_{\mathrm{subjet}}^{\mathrm{max}}}

\newcommand{\Wp}  {W^{\prime}}
\newcommand{\Up}  {U(1)^{\prime}}
\newcommand{\Zp}  {Z^{\prime}}
\newcommand{\ZpB} {\Zp_{B}}
\newcommand{\gB}  {g_{B}\ds}

\begin{document}

\title{$\mu_x$ boosted-bottom-jet tagging and $\Zp$ boson searches}

\author{Keith~Pedersen}
\email{kpeders1@hawk.iit.edu}
\affiliation{Department of Physics, Illinois Institute of Technology, Chicago, Illinois 60616-3793, USA}

\author{Zack~Sullivan}
\email{Zack.Sullivan@IIT.edu}
\affiliation{Department of Physics, Illinois Institute of Technology, Chicago, Illinois 60616-3793, USA}

\preprint{IIT-CAPP-15-05}

\date{November 18, 2015}

\pacs{13.20.He,13.87.Fh,14.70.Pw,12.60.Cn}

\begin{abstract}
  We present a new technique for tagging heavy-flavor jets with $p_T >
  500$ GeV called ``$\mu_x$ tagging.''  Current track-based methods of
  $b$-jet tagging lose efficiency and experience a large rise in fake
  rate in the boosted regime.  Using muons from $B$ hadron decay, we
  combine angular information and jet substructure to tag $b$ jets,
  $c$ jets, light jets, and ``light-heavy'' jets (those containing $B$
  hadrons from gluon splitting).  We find tagging efficiencies of
  $\epsilon_b = 14\%$, $\epsilon_c = 6.5\%$,
  $\epsilon_{\mathrm{light-light}} = 0.14\%$, and
  $\epsilon_{\mathrm{light-heavy}} = 0.5\%$, respectively, that are
  nearly independent of transverse momentum at high energy.  We
  demonstrate the usefulness of this new scheme by examining the
  discovery potential for multi-TeV leptophobic $Z^\prime$ bosons in
  the boosted-$b$-tagged dijet channel at the Large Hadron Collider.
\end{abstract}

\maketitle


\section{Introduction}

Searches for new narrow massive vector current particles, generally
called $\Zp$ or $\Wp$ bosons, are a main focus of the exotics groups in 
experiments at the Large Hadron Collider (LHC). These particles arise in 
many extensions of the standard model (SM), such as the sequential 
standard model~\cite{Altarelli:1989ff}, broken $\SUtwo[L] \times \SUtwo[R]$ 
symmetry~\cite{Mohapatra:1974hk,Mohapatra:1974gc,Senjanovic:1975rk}, 
grand unified models~\cite{Cvetic:1983su,Hewett:1988xc,Leike:1998wr},
Kaluza-Klein excitations in models of extra dimensions~\cite{Randall:1999ee,Randall:1999vf}, 
non-commuting extended technicolor~\cite{Chivukula:1995gu}, general extended 
symmetries~\cite{Dobrescu:leptophobic-Z', Dobrescu:leptophobic-anamalons}, 
and more.

Using 8~TeV LHC data, the ATLAS~\cite{Aad:2014cka} and
CMS~\cite{Khachatryan:2014fba} collaborations set bounds on many types
of $\Zp$ bosons that decay to dileptons below around 2.9~TeV. A more
challenging search is for leptophobic gauge bosons, such as a
top-color $\Zp$ boson, which is excluded up to
$2.4$~TeV~\cite{Chatrchyan:2013lca,Aad:2015fna,Khachatryan:2015sma},
or a right-handed $\Wp$ boson, which is excluded for SM-like couplings
up to $1.9$~TeV~\cite{Aad:2014xea,Chatrchyan:2012gqa}.  This latter
boson is most strongly constrained by the $\Wp \to tb$ final
state~\cite{Sullivan:Wprime,Duffty:boostedBtag}.

Flavor tagging such states becomes challenging in searches for vector
boson resonances above 1.5~TeV, where dijet signals contain
boosted-top jets~\cite{Kaplan:2008ie,CMS:2009lxa,Almeida:2008tp,
Thaler:2010tr,Anders:2013oga,CMS:2014fya,Almeida:2015jua,Kasieczka:2015jma,
Usai:2015vva} and
boosted-bottom jets~\cite{Duffty:boostedBtag,CMS:2013vea,ATLAS:bTagging-in-dense-env}.
For example, the systematic uncertainties in $b$-tagging efficiency and
fake rates dominate the current $\Wp \to tb$ limits, and have so far closed the
$\Zp \to \pair{b}$ searches from consideration. This is evident in the
ATLAS $\Wp$ searches~\cite{ATLAS:W'-search-8TeV, Aad:2014xra}, which found a 35\%
uncertainty in the $b$-jet tagging efficiency for jets with $\pT$
above 500 GeV (i.e. $M_{\Wp}\apprge$ 1 TeV). This is mainly driven by
a lack of clean samples of high-$\pT$ $b$~jets tagged with a
complementary method, which are necessary to cross-check the
signal/background efficiencies of the
$b$~tags~\cite{ATLAS:bTag-calibrate-ptRel, ATLAS:bTag-calibrate-tt,
  ATLAS:bTag-calibrate-D*}.  Most concerning is the dramatic rise of
the $b$-tagging fake rate for jets initiated by light quarks as jet
transverse momentum $\pT \to \bigO{\TeV}$ \cite{ATLAS:bTagging-dense}.
For instance, a CMS search for exotic resonances above 1.2~TeV
encountered fake rates above 10\% per
jet~\cite{CMS:quantum-black-holes}.

This paper proposes an improvement to the \emph{boosted-bottom-jet tag}
first proposed in Ref.~\cite{Duffty:boostedBtag}.
Here, the focus is on $b$ quarks which are themselves highly boosted,
instead of boosted topologies which contain bottom quarks
(e.g., boosted $t\to W b$ or $H\to \pair{b}$).
In Sec.~\ref{sec:tagging} we explain why existing tagging methods are 
insufficient at high energies, and then derive from first principles a 
muon-based tag we call {\new} boosted-bottom-jet tagging.
In Sec.~\ref{sec:tagging-results} we present the {\new} tagging efficiencies
for bottom and charm flavored jets, along with small light-jet fake rates,
using a detailed simulation based on the ATLAS detector.

In order to determine the efficacy of this {\new} tag for new physics
searches, we perform a full signal and background study for a
leptophobic $\Zp$ boson
\cite{Dobrescu:leptophobic-Z',Dobrescu:leptophobic-anamalons}.  This
model assumes a flavor-independent $\ZpB$ gauge coupling to SM quarks
\begin{equation}
\lagrange=\frac{\gB}{6}Z_{B\mu}^{\prime}\bar{q}\gamma^{\mu}q ,
\end{equation}
and demonstrates the power of our new boosted-bottom tag. We conclude
in Sec.~\ref{sec:fin} with a discussion of other searches for physics
beyond the standard model that this tag enables, and experimental
information that could further improve the fake rejection for our
algorithm.


\section{Tagging a heavy-flavored jet \label{sec:tagging}}

Heavy-quark ($b$ or $c$) initiated jets shower and hadronize in a
manner that is distinct from light parton ($d$, $u$, $s$, or $g$)
initiated jets. The large masses of the heavy quarks
($m\apprge\lambdaQCD$) cause their fragmentation functions to peak
near $z=1$. Thus, $b$ and $c$~quarks tend to retain their momentum
during fragmentation~\cite{Cacciari:heavy-quark-PFF}, spawning heavy
hadrons which carry a large fraction of their jet's momentum.  These
hadrons have long lifetimes
($c\tau(B/D)\approx\bigO{\powerUnit{-4}{m}}$), and the decay daughters
of even moderately boosted $b/c$~hadrons will point back to secondary
vertices (SV) whose impact parameters (IP) are far enough from the
primary vertex to be resolved, but close enough to distinguish them
from other meta-stable particles (e.g.\
$c\tau(\Kshort)=3\times\powerUnit{-1}{m}$). Additionally, the
significant rate of semi-leptonic decay of $b/c$~hadrons
($\mathcal{B}(X_{b/c}\to l\,\nu_{l}\,Y)\approx0.1$ for each
$l\in\{e,\mu\}$) enriches their jets with energetic leptons.  Since
bottom hadrons decay primarily to charm hadrons, $b$~jets have twice
the probability of $c$ jets to contain leptons.


\subsection{Challenges for existing $b$ tags}

Modern $b$-tagging algorithms are essentially track-based tags that
search for evidence of a secondary
vertex~\cite{ATLAS:bTag-performance-7TeV,CMS:bTag-performance-7TeV}.
While they frequently use neural nets and multiple inputs, their
efficiencies are predominantly determined by the impact parameter of a
jet's tracks and the mass of its reconstructed SV.  Although light
jets also contain secondary vertices (e.g. $\Kshort/\Lambda$ decay or
material interaction \cite{ATLAS:material-SV}), this background is
largely reducible for jets with $\pT<300$~GeV, giving track tags high
$b$~jet efficiency (50-80\%) and light jet fake rates of $\bigO{1\%}$.
Above $\pT=300$~GeV, the increasingly boosted nature of the jet makes
track-tagging difficult. Boosted tracks bend less, and are thus harder
to constrain and more sensitive to tracker resolution and alignment.

These problems are exacerbated in boosted heavy-flavor jets, where the
primary hadron can decay \emph{after} traversing one or more pixel
layers, making it difficult or impossible for its daughters to produce
the ``high purity'' tracks needed by most SV tagging algorithms.
Additionally, if these collimated daughters strike \emph{adjacent}
pixels, they can create a ``merged cluster'' which also hinders
reconstruction \cite{CMS:bTag-performance-7TeV,ATLAS:merged-clusters}.
These problems are well exemplified by Fig.~12 of
Ref.~\cite{CMS:bTag-simulation}, where the light-jet fake rate of the
``track counting high purity'' algorithm increases 100-fold as jet
$\pT$ increases from 100~GeV to 1~TeV.

Another component of current $b$-tagging algorithms is $\pTrel$
tagging, which measures the momentum of leptons transverse to the
centroid of their jet.  Compared to light hadrons, heavy hadrons have
a larger mass and carry a larger fraction of their jet's momentum;
thus, leptons produced by heavy hadrons will have more energy and will
arrive at wider angles inside the jet.  These effects conspire to
produce larger values of $\pTrel$~\cite{ATLAS:soft-muon-tagging,CMS:bTag-simulation}.  Since electrons are difficult to identify
inside jets, $\pTrel$ tagging generally utilizes only muons. In ATLAS
and CMS, muon $\pTrel$ tags give $\about{10}\%$ $b$~jet efficiency and
a light jet rejection (inverse tagging efficiency) of about 300
\cite{ATLAS:soft-muon-tagging}. However, once jet $\pT$ exceeds about
140~GeV, the underlying boost makes $\pTrel$ distributions for heavy
and light jets nearly
indistinguishable~\cite{ATLAS:bTag-calibrate-ptRel}, precluding the
tag.


\subsection{The {\new} boosted-$b$ tag}

The failure of existing tagging methods to adequately reject
high-$\pT$ light jets is a problem. For track tagging, it is
essentially a problem of detector resolution, so any improvements will
likely involve novel utilization of the hardware and track
observables.  For $\pTrel$ tagging it is potentially a problem of
definition; $\pTrel$ dilutes a well measured muon angle with a more
poorly measured muon energy. This drove the development of the
``boosted-bottom tag''~\cite{Duffty:boostedBtag}, a purely angular tag
on jets containing muons within $\Delta R=0.1$ of their centroid.
This tag achieves nearly ideal signal efficiency (given the muonic
branching fraction), but suffers from a continuous rise with energy in
light jet fake rate. Since the centroid of an \emph{entire} jet is not
necessarily aligned with the $B$~hadron's decay, and the boost cone of
muon emission should tighten as the boost increases, $b$~jet decay
should be reexamined in the context of jet substructure.  This will
provide the basis for a new heavy-flavor tag, which we dub the
``{\new} boosted-$b$ tag.''


\subsubsection{Theory of the {\new} tag}

\begin{figure}[htb]
\includegraphics[width=\columnwidth]{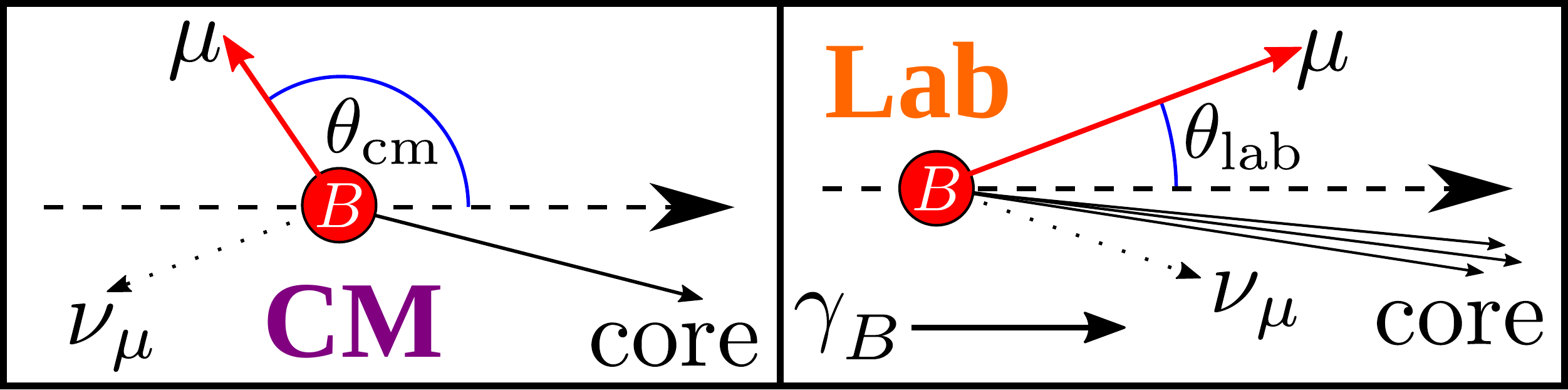}
\caption{Nomenclature for the center-of-momentum frame and boosted lab
frame. \label{fig:Boosted-Nomenclature}}
\end{figure}

Consider a jet containing a $B$~meson that decays semi-muonically.  In
the decay's center-of-momentum (CM) frame, the muon is emitted with
speed $\betaMuCM$ and angle $\thetaCM$ with respect to the boost
axis (see Fig.~\ref{fig:Boosted-Nomenclature}).  In the lab frame, the
$B$~meson's decay products are boosted (by $\gammaB$) into a
\emph{subjet} with a hadronic ``core'' (which is typically a charm
hadron) with four-momentum
\begin{equation}
    \pSubjet=\pCore+\pMu+\pNu,  \label{eq:pSubjet}
\end{equation}
and the muon now makes the angle $\thetaLab$ with the $B$ meson's direction.

Using basic kinematics, we can define a lab frame observable
\begin{equation}
    x\equiv\gammaB\,\tan(\thetaLab)=\frac{\sin(\thetaCM)}{\kappa+\cos(\thetaCM)} ,  \label{eq: x definition}
\end{equation}
where $\kappa\equiv\beta_{B}/\betaMuCM$.  While $\kappa$ depends on
the boost of the muon in the CM frame --- which is generally not
measurable --- $x$ itself has almost no dependence on $\kappa$ when
the system is sufficiently boosted ($\gammaB\gg\gammaMuCM\apprge 3$),
as $\kappa\to 1$ in this limit. Fortuitously, the kinematics of both
$B$~meson decay and the jets of interest (jet $\pT>\minJetPt$ GeV)
ensure this condition, giving lab frame muons from boosted $B$~meson
decay a nearly universal $x$ distribution.

Assuming isotropic CM emission $\left(dN/d\Omega=\frac{1}{4\pi}\right)$,
the differential muon count $N$ is
\begin{equation}
\frac{dN}{d\thetaCM}=\frac{1}{2}\sin(\thetaCM).\label{eq: CM isotropic}
\end{equation}
When $\kappa \ge 1$, $dN/dx$ can be written in the lab frame as
\begin{equation}
\frac{dN}{dx}=\frac{2x}{(x^{2}+1)^{2}}\,K(x,\kappa) ,    \label{eq:dOmega_dx}
\end{equation}
where
\begin{equation}
K(x,\kappa)=\begin{cases}
\frac{(1+\kappa^{2})+x^{2}(1-\kappa^{2})}{2\sqrt{1+x^{2}(1-\kappa^{2})}} & 0\le x\le1/\sqrt{\kappa^{2}-1}\\
0 & \textrm{everywhere else}
\end{cases}.
\end{equation}
Here, $K(x,\kappa)$ enforces the boundary of the boost cone; i.e., 
when $\gammaB\gg\gammaMuCM$, the maximum value of $x$ which a lab
frame muon can achieve is $x=\sqrt{\gamma_{\mu,\mathrm{cm}}^2-1}$.

\begin{figure}[htb]
\includegraphics[width=0.5\columnwidth]{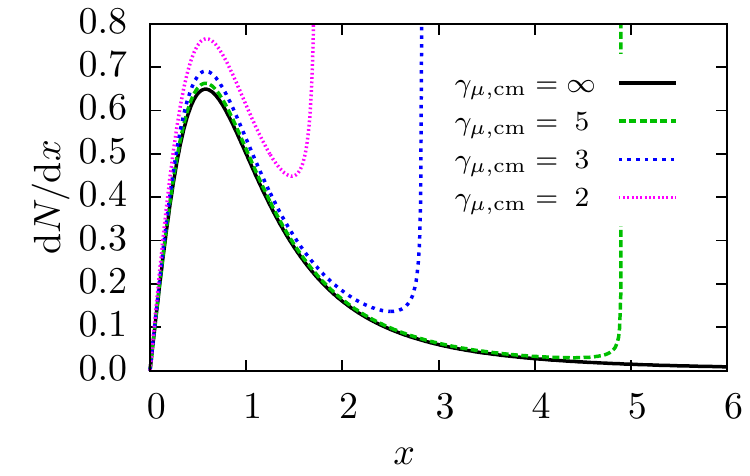}
\caption{Theoretical muon distribution $dN/dx$ vs.\ $x$ in the lab frame
(with $\beta_{B}\to 1$) for various $\gammaMuCM$.
\label{fig:dNdx}}
\end{figure}

As $K(x,\kappa) \to 1$ in the boosted limit $(\kappa=1)$, the muon
distribution in $x$ approaches a universal shape $2x/(x^2+1)^2$. 
We see the approach to this limit in Fig.~\ref{fig:dNdx}, where the
distribution of muon count vs.\ $x$ is shown for several muon boosts
$\gammaMuCM$. Since the typical muon boost in the CM frame $\gammaMuCM>3$, 
where deviations from the universal shape are small, 
the $dN/dx$ of a typical muon is well represented by the boosted limit.
This makes this universal shape useful for identifying 
muons from a boosted decay.

Assuming that all muons follow the universal shape, 
we calculate the largest value of $x$ which confines a
fraction $\rho$ of lab frame muons (i.e. the inverse cumulative
distribution),
\begin{equation}
x_{\rho}=\sqrt{\frac{\rho}{1-\rho}}.
\end{equation}
We define a cut $\xCut = x_{90\%}$ which accepts 90\% of muons
compatible with boosted $B$ hadron decay. In addition, we use the hard
fragmentation of $b$ quarks to motivate a cut on the $\pT$ fraction of
the $B$ hadron subjet to the total jet $\pT$
\begin{equation}
\fSubjet\equiv\frac{\pTrm{subjet}}{\pTrm{jet}} \ge 0.5.
\end{equation}
These two cuts ($x\le\xCut$ and $\fSubjet\ge\fCut$) define the {\new}
boosted-bottom-jet tag.


\subsubsection{Reconstructing $\pSubjet$ and measuring $x$}

Although $x$ is defined in terms of an isolated decay of a bottom
hadron, $\pSubjet$ will overlap other energy in the jet. Furthermore,
half of a $b$~jet's semi-muonic decays come from charm
hadrons. Therefore, it is not possible to measure $\gammaB$ --- only
$\gammaSubjet$. In spite of this limitation, we will see that it is
still possible to reconstruct a meaningful $x$.

First, jets are clustered using the {\antiKt} algorithm and a radius
parameter $R=0.4$. Muons are allowed to participate in jet clustering,
which lets hard muons seed jet formation. Candidates for {\new}
tagging must contain a \emph{taggable} muon ($\pTmu\ge\minMuPt$) to
ensure good muon reconstruction. While a taggable muon's associated
neutrino is inevitably lost, most of the muon and neutrino momentum
comes from their shared boost, making the muon an acceptable neutrino
analog.  We use the simplest choice: $\pNu=\pMu$.

A jet's internal list of candidate cores is obtained by reclustering
the jet with the {\antiKt} algorithm using $\Rcore=\coreR$; this
radius is designed to localize the core to a $3\times3$ grid, based on
the fixed width $w$ of the calorimeter towers ($\sqrt{2}w<\Rcore<2w$).
All jet constituents are used during reclustering (allowing taggable
muons to seed core formation) \emph{except} towers failing a cut on
jet $\pT$ fraction (we choose $\pTfrac{tower}{min}=\minTowerf$); this
reduces the core's sensitivity to pileup, the underlying event, and
soft QCD.  Since the calorimeter granularity produces an ill-measured
core mass, we fix the mass of each core candidate to a charm hadron
mass $\mCore=\coreMass$.  We identify the ``correct'' core as the
candidate which brings $\sqrt{\pSubjet[2]}$ closest to $\mB$, the
nominal mass of the $b$~hadron admixture (we choose $\mB=\BMass$).

\begin{figure}[htb]
\includegraphics[width=0.5\columnwidth]{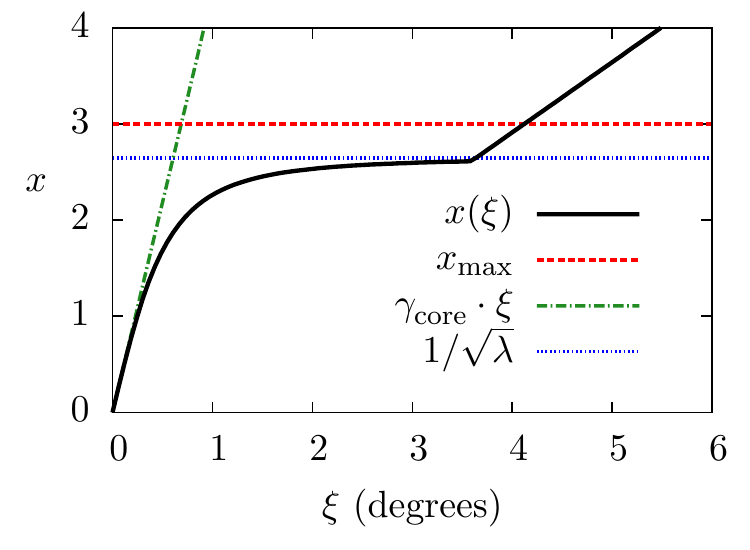}
\caption{$x(\xi)$ for a subjet with $\gammaCore=250$ and a hard muon $(\lambda=1/7)$. \label{fig:xBehavior}}
\end{figure}

Given our neutrino strategy ($\pNu=\pMu$), we can study the value of
$x$ that will be observed for an \emph{arbitrary} muon-associated
subjet (which \emph{could} be the remnants of a $B$ hadron, but could
also be a random association of jet constituents). Such a subjet can
be fully described using three lab frame observables: $\gammaCore$
(the energy of the core), $\lambda={2E_{\mu}}/{E_{\mathrm{core}}}$
(the energy of the muon, relative to the core), and $\xi$ (the
lab-frame angle between the muon and the \emph{core}). Assuming that
both the muon and the core are ultra-relativistic in the lab frame
(i.e. $\beta\to 1$),
\begin{equation}
x(\xi) \approx \underset{\gammaSubjet}{\underbrace{\gammaCore \frac{1+\lambda}{\sqrt{1+2\lambda\,\gammaCore[2] [1-\cos(\xi)]}}}} \; \underset{\tan(\thetaLab)}{\underbrace{\frac{\sin(\xi)}{\cos(\xi)+\lambda}}},
\end{equation}
where the square root term scales $\mCore$ to the larger $\mSubjet$.
This form reveals two distinct $\xi$ regimes, visible in
Fig.~\ref{fig:xBehavior}.  When $\xi$ is vanishingly small,
$x(\xi)\approx\gammaCore\cdot\xi$.  For intermediate $\xi$ (large
enough to dominate $\mSubjet$, but small enough that
$\tan(\thetaLab)\approx\frac{\xi}{1+\lambda}$), $x(\xi)$ flattens into
a plateau at $x\approx1/\sqrt{\lambda}$.

The $x$ plateau exists because, as $\xi$ rises, every increase in
$\tan(\thetaLab)$ is balanced by an increase in $\mSubjet$.  Once
$\mSubjet\gg\mB$, the reconstructed subjet is no longer consistent
with a $B$~hadron decay.  This requires limiting $\mSubjet$ (we choose
$\mSubjetMax=\maxSubjetMass$), which forces the $x$ of poorly
reconstructed (or fake) subjets to abruptly return to a nearly linear
$\xi$ dependence. This discontinuity is visible in
Fig.~\ref{fig:xBehavior}.  When the plateau is below $\xCut$, the
maximum taggable $\xi$ is
\begin{equation}
\xiCut[\mathrm{hard}] \approx \frac{\xCut}{\gammaCore}\left(\frac{\mSubjetMax}{\mCore}\right).
\end{equation}
Thus, for \emph{hard} muons ($\lambda\ge\xCut[-2]$), $\xCut$ is a
purely angular cut which scales inversely proportional to the energy
of the core, with no additional dependence on the energy of the
muon. On the other hand, when the plateau is above $\xCut$,
\begin{equation}
\xiCut[\mathrm{soft}] \approx\frac{\xCut}{\gammaCore}\left(\frac{1}{\sqrt{1-\lambda\,\xCut[2]}}\right).\label{eq:phiMax-softMuon}
\end{equation}
So for \emph{soft} muons ($\lambda<\xCut[-2]$), $\xCut$ is a much
tighter angular cut which scales with the energy of both the core and
the muon.  But unless $\lambda$ is near $\xCut[-2]$ (the soft/hard
muon boundary), $\xiCut[\mathrm{soft}]$ is only mildly sensitive to
$\lambda$.

This means that $\xCut$ is effectively a dual angular cut: a very
tight cut for soft muons, a looser cut for hard muons, and a quick
transition region (as a function of muon energy) between the two
classes.  Thus, the {\new} tag depends primarily on well measured
angles.  For convenience, we summarize the parameters chosen for
{\new} tagging in Table \ref{tab:tagging-parameters.}.

\begingroup
\begin{table}[!htb]
\caption{A summary of parameters chosen for {\new} boosted 
bottom jet tagging.\label{tab:tagging-parameters.}}
\begin{ruledtabular}
\begin{tabular}{cc|cc|cc}
$\boldsymbol{R}$ & {\jetR} & 
$\boldsymbol{m}_{\mathbf{core}\ds}$ & $\coreMass$ & 
$\boldsymbol{p_{T, \mu}^{\mathbf{min}}}$ & $\minMuPt$ \tabularnewline
$\boldsymbol{R}_{\mathbf{core}}$ & {\coreR} &
$\boldsymbol{m_{B}\ds}$ & $\BMass$ &
$\boldsymbol{x}_{\mathbf{max}}$ & {\cutX} $(x_{\cutXrho}\ds)$\tabularnewline
$\boldsymbol{f}_{\mathbf{tower}}^{\mathbf{min}}$ & $\minTowerf$ &
$\boldsymbol{m}_{\mathbf{subjet}}^{\mathbf{max}}$ & $\maxSubjetMass$ &
$\boldsymbol{f}_{\mathbf{subjet}}^{\mathbf{min}}$ & {\minSubjetF} \tabularnewline
\end{tabular}
\end{ruledtabular}
\end{table}
\endgroup


\section{{\new} tagging results\label{sec:tagging-results}}


We extract the {\new} tagging efficiency for individual jets by
simulating detector reconstruction for samples of flavored dijets.  We
generate all samples at $\sqrtS$ using {\mgCite} with {\pdfSetCite}.
We use {\pythiaCite} for all fragmentation, hadronization, and decay,
using the default {\pythia} tune and PDF set for everything except
pileup, for which we use the settings described in Table~7 of
Ref.~\cite{ATLAS:pileup-pythia}.  To allow in-flight muon production,
we activate $\Klong$, $K^{+}$ and $\pi^{+}$ decays.

We use {\fastjetCite} to reconstruct jets, and a modified version of
{\delphesCite} to simulate the ATLAS detector at the LHC. Since the
{\new} tag relies heavily on muon angle, with in-flight
$\pi^{+}/K^{+}$ decays being a large source of muon background, we
developed a custom module \texttt{AllParticlePropagator} to properly
handle such decays. The module which implements {\new} tagging
\texttt{MuXboostedBTagging} (available on GitHub
\cite{Delphes:myDelphes}) can be used in conjunction with {\delphes}'
default \emph{b} tagging module \texttt{BTagging}.  It important to
note that, until the most recent version of {\delphes} (3.3), the
default {\delphes} cards define \texttt{BTagging} efficiencies which
are \emph{not accurate} at high $\pT$ (e.g. light-jet fake rates are
constant everywhere, and $b/c$~jet efficiencies are constant for jets
with $\pT\apprge$ 150~GeV).  The {\delphes} 3.3 efficiencies for 1--2
TeV jets are now 14--28\% for $b$-tags and 1--2\% for light jet fake
rates.  Our goal is to provide similar $b$-tagging efficiency with a
factor of 10 improvement in fake rates.

Muon reconstruction efficiencies and $\pT$ resolutions are taken from
public ATLAS plots \cite{ATLAS:detector-performance-2008,
  ATLAS:public-muon-plots} for \emph{standalone} muons (muons seen in
the Muon Spectrometer (MS), but not necessarily the main
tracker). Because the MS experiences limited punch-through, it can
reconstruct muons with $\pT\ge10$~GeV with high efficiency (95--99\%),
even inside boosted jets.  Because we focus on the ATLAS MS, our
results reflect the holes for detector services and support feet,
which cause (i) a dip in muon reconstruction efficiency at
$\eta=0$~\cite{ATLAS:muon-performance-run1}, precisely where the dijet
$dN/d\eta$ distribution peaks, and (ii) 80\% geometric acceptance of
the Level-1 muon trigger in the barrel~\cite{ATLAS:muon-trigger}.
This latter restriction can be resolved by relying on jet triggers
(jet $\pT$ or event $\HT$) to select pertinent events, since {\new}
tagging only works for high-$\pT$ jets.

There are several sources of standalone muon background which we are
unable to simulate: (i)~cosmic muons, (ii)~decay muons from particles
produced in the calorimeter shower, (iii)~fake muons from
punch-through, and (iv)~fake muons from noise. Nonetheless, since the
{\new} tag is effectively a tight angular cut with a reasonably high
$\pTmu$ threshold, we expect these backgrounds to be negligible
compared to the light jet background which we simulate.

The direction of the core is extremely important in {\new} tagging, and
tracks would provide the best information. However, the core's
intrinsic collimation hampers track reconstruction in a manner
difficult to model in a fast detector simulation. As such, we build
jets (and cores) solely from calorimeter towers and muons. The coarse
granularity of the hadronic calorimeter (HCal) is mitigated by using
the finer granularity of the EM calorimeter (ECal) to orient the
combined tower (``ECal pointing'').  This is implemented in
{\delphes}' \texttt{Calorimeter} module by giving both ECal and HCal
the segmentation of ATLAS's ECal Layer-2
($\Delta\phi\times\Delta\eta=0.025\times0.025$ in the barrel). 
To ensure that we are not overly sensitive to this resolution, we also
test a granularity twice as coarse $(0.05\times0.05)$, finding negligible
degradation in the heavy jet tagging efficiency, with only a slight
rise in light jet fake rate (1.2 times larger at $\pT=600$~GeV, 
but dropping to no increase at 2.1~TeV).


\subsection{Tagging efficiencies}
\label{subsec:eff}

To test the {\new} tag, we create ten 200 GeV wide samples of
$\pair{b}$, $\pair{c}$, and $\pair{j}$ ($j\in\{u,d,s,g\}$) spanning
$\pT=$ 0.1--2.1 TeV. We then find the efficiency to tag the top two
jets (ranked by $\pT$) in each event.  Since heavy hadrons from gluon
splitting ($g\to b\bar b/c\bar c$) are an inevitable component of our
light-jet sample, especially at high $\pT$, it is important to
determine the extent to which this background can be reduced. We sort
the light jet sample via the truth-level flavor of a taggable muon's
primary hadronic precursor.  This classifies each attempted tag as
\emph{light-heavy} (where the muon descends from a $b/c$~hadron inside 
a jet initiated by a light parton) or \emph{light-light} (where the muon's
lineage is a purely light-flavored).

\begin{figure*}[htb]
\subfloat[\label{fig:tagging-efficiency-vs-pT}]{\includegraphics[width=0.5\textwidth]{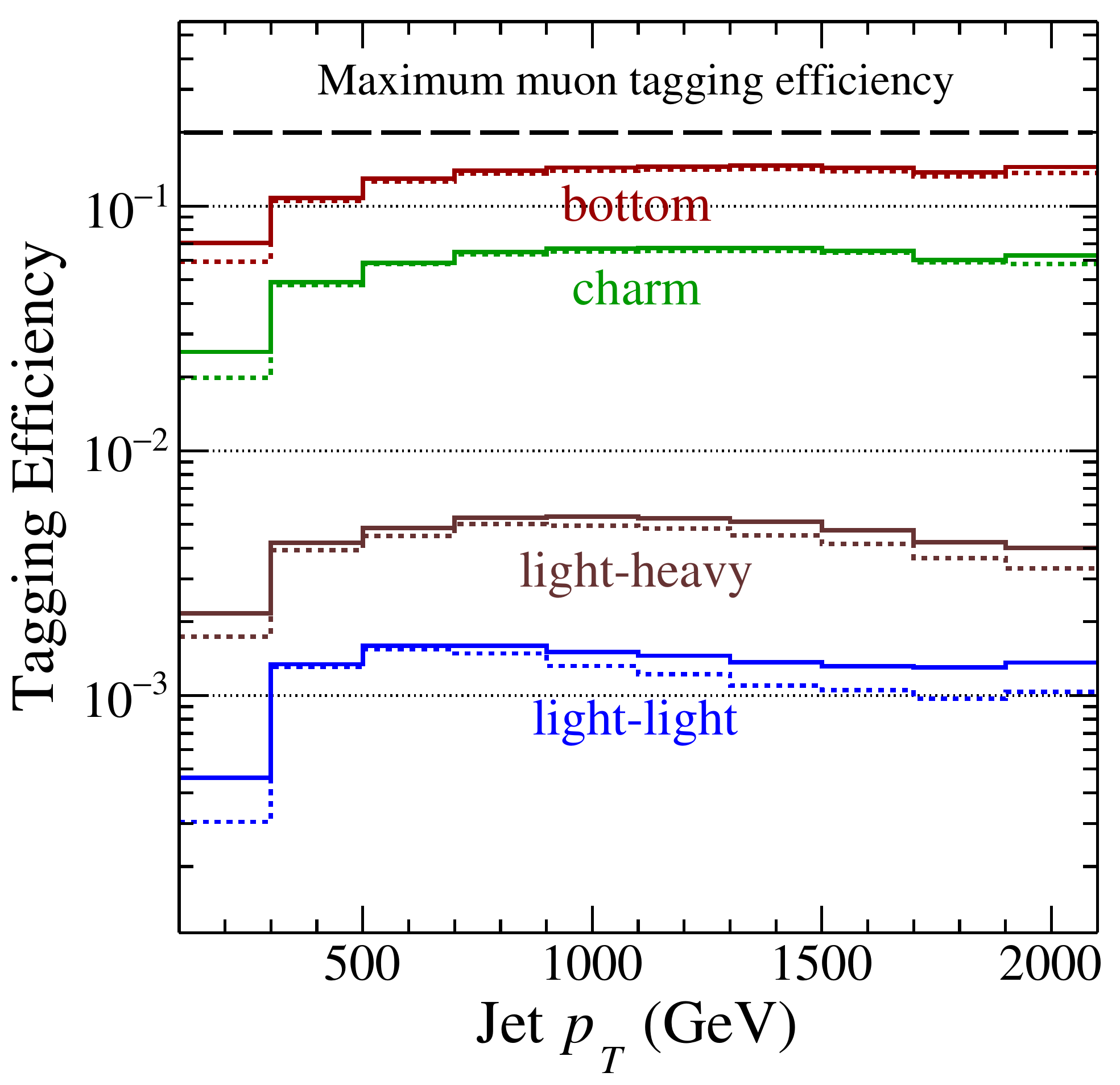}}
\subfloat[\label{fig:tagging-efficiency-vs-eta}]{\includegraphics[width=0.5\textwidth]{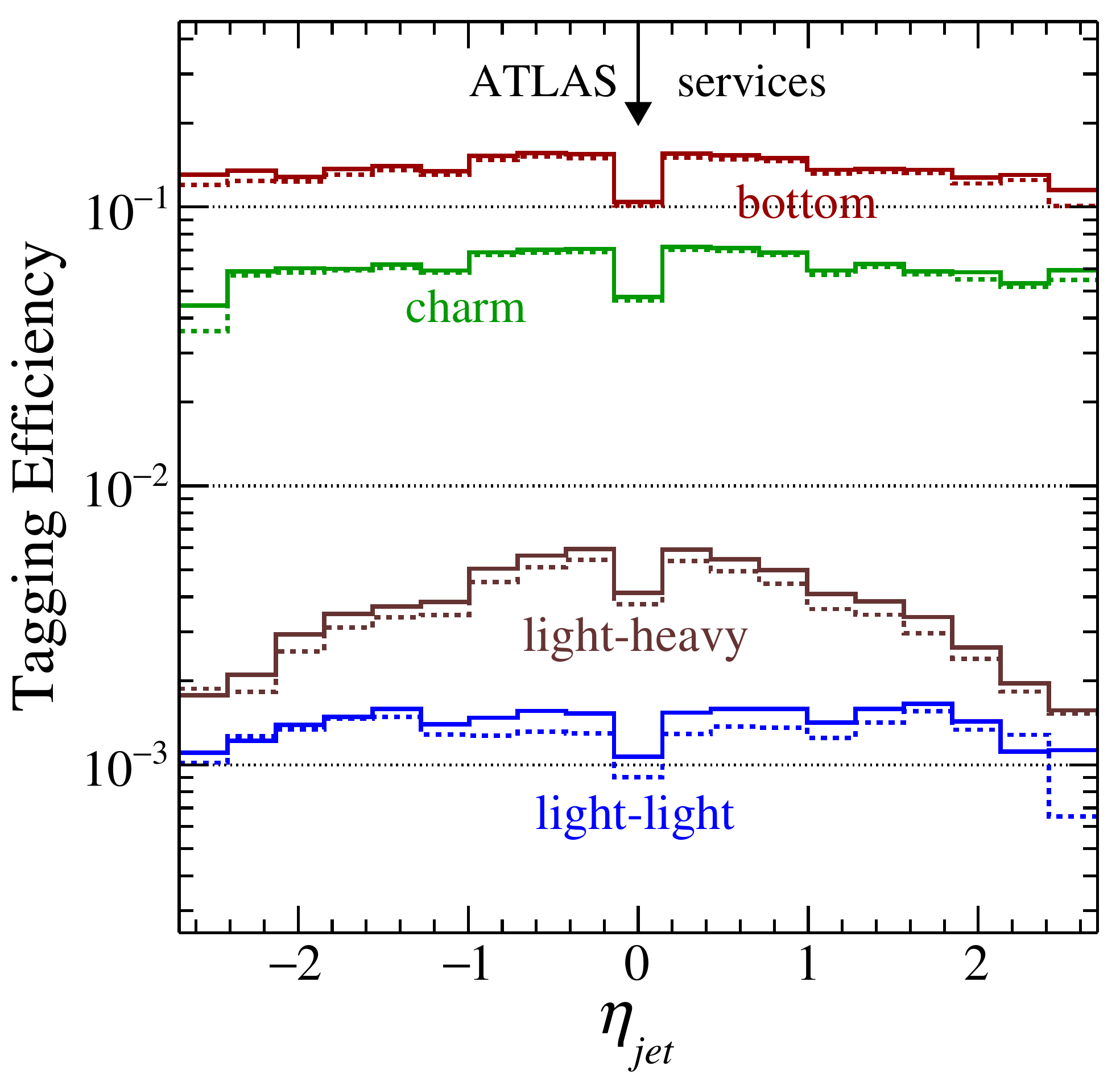}}
\caption{{\new} tagging efficiency vs. (a) jet $\pT$ and (b) $\etaJet$.
\label{fig:tagging-efficiency}}
\end{figure*}

In Fig.\ \ref{fig:tagging-efficiency} we show our predicted
efficiencies for the four classes of {\new} tags.  The solid lines
represent the efficiencies without pileup, while the dotted lines show
the efficiencies when a random number of pileup events (drawn from a
Poisson distribution with $\mu=\pileupMu$) are added to each hard
event. Since we do not utilize non-muon tracking, and are working with
TeV-scale jets, we do not attempt any pileup subtraction.

Each $\pT$ bin in Fig.~\ref{fig:tagging-efficiency-vs-pT} sums over
all available $\etaJet$. When the boosted approximations are valid
(jet $\pT\ge\minJetPt$~GeV), the efficiency to tag heavy jets is
nearly flat versus $\pT$, while the efficiency to tag light jets
decreases slightly. We find asymptotic tagging efficiencies of
$\epsilon_b = 14\%$, $\epsilon_c = 6.5\%$,
$\epsilon_{\mathrm{light-light}} = 0.14\%$, and
$\epsilon_{\mathrm{light-heavy}} = 0.5\%$, respectively.  This
light-light rejection provides us the full factor of 10 improvement
over existing algorithms.  At low-$\pT$ (where $B$ hadrons are no
longer strongly boosted and track tagging is superior) all {\new}
efficiencies plummet, although the relative rates remain approximately
the same. Notice that pileup actually \emph{improves} the performance
of {\new} tagging above 1 TeV, causing almost no degradation in
heavy-jet efficiencies, but a significant drop in light-jet
efficiency.  This is a consequence of the increased probability for
light jets to reconstruct a subjet with a low fraction of total jet
energy, thereby failing the cut on $\fSubjet$.

Since the {\new} tag is not effective at low $\pT$, each $\etaJet$ bin
in Fig.~\ref{fig:tagging-efficiency-vs-eta} requires $\pT\ge\minJetPt$
GeV.  We can see that both heavy and light-light jet efficiencies are
flat with $\etaJet$.  The light-heavy efficiency decreases
significantly with $|\etaJet|$, indicating a rising rejection of heavy
hadron background from gluon splitting.  This suggests the intriguing
possibility that the $g\to b\bar b$ contribution to $b$ jets could be
extracted from data, and used to calibrate the Monte Carlo event
generators for highly boosted jets.

\begin{figure*}[htb]
\subfloat[]{\includegraphics[width=0.5\textwidth]{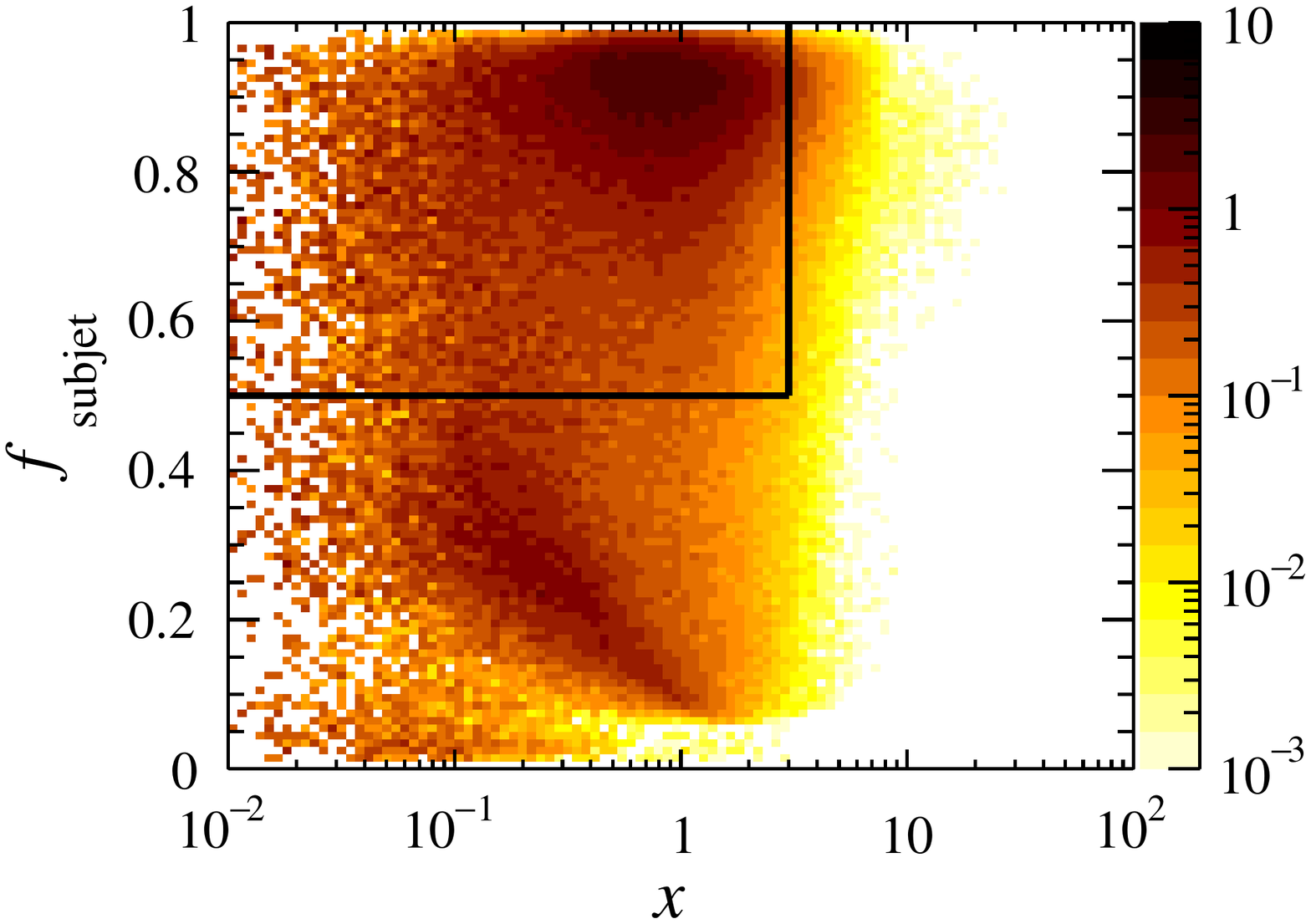}}
\subfloat[\label{fig:charmXF}]{\includegraphics[width=0.5\textwidth]{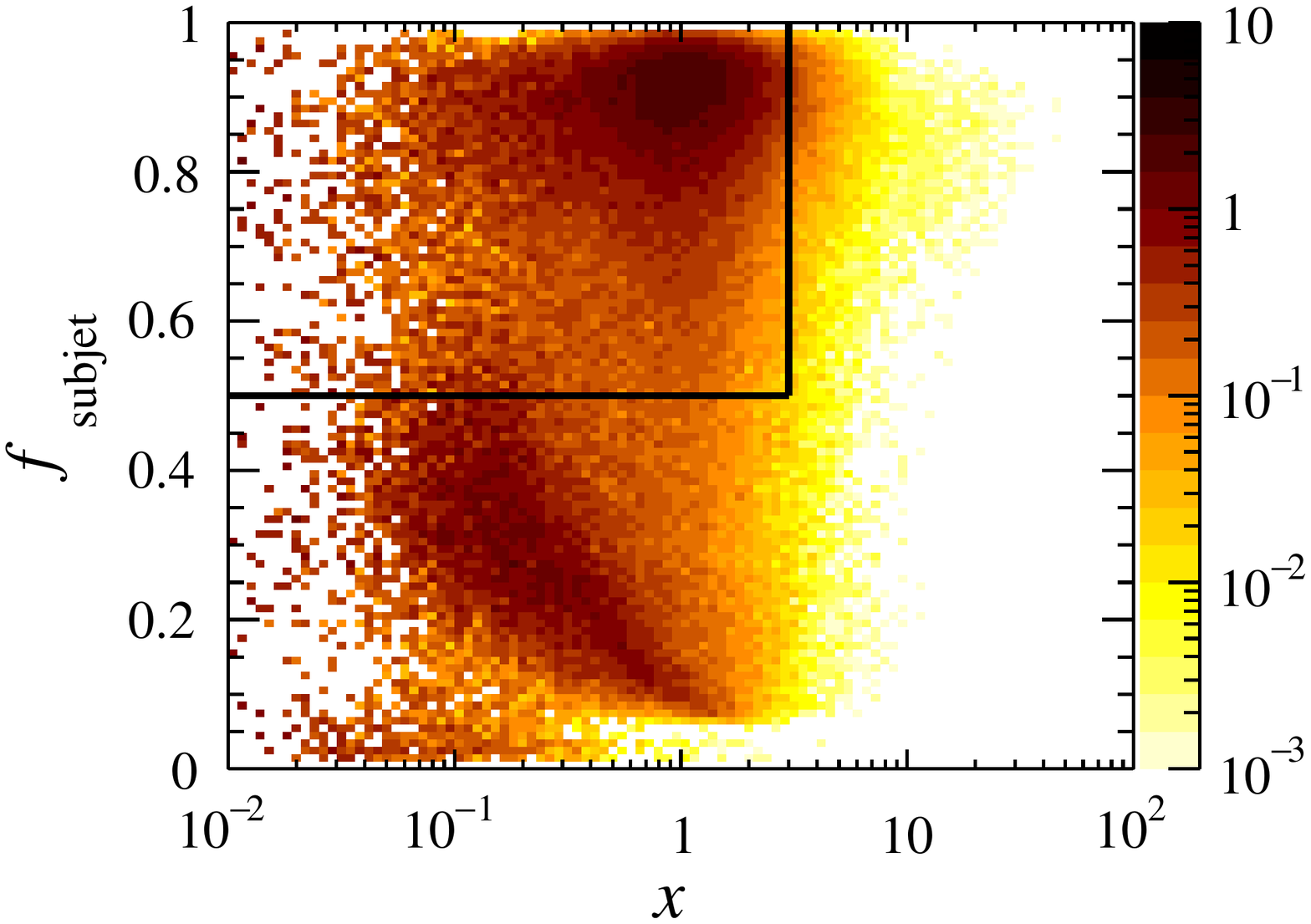}}\\
\subfloat[]{\includegraphics[width=0.5\textwidth]{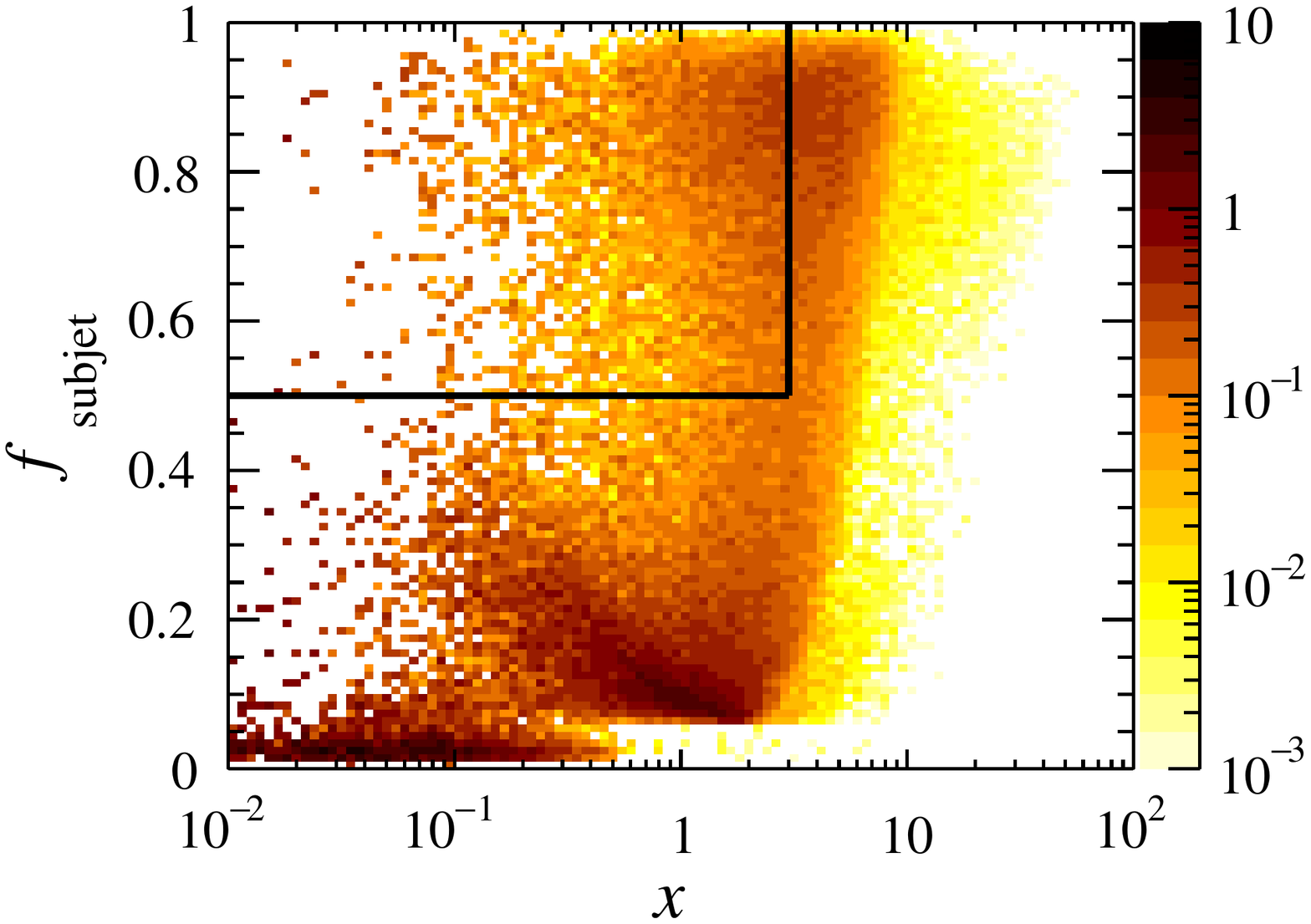}}
\subfloat[]{\includegraphics[width=0.5\textwidth]{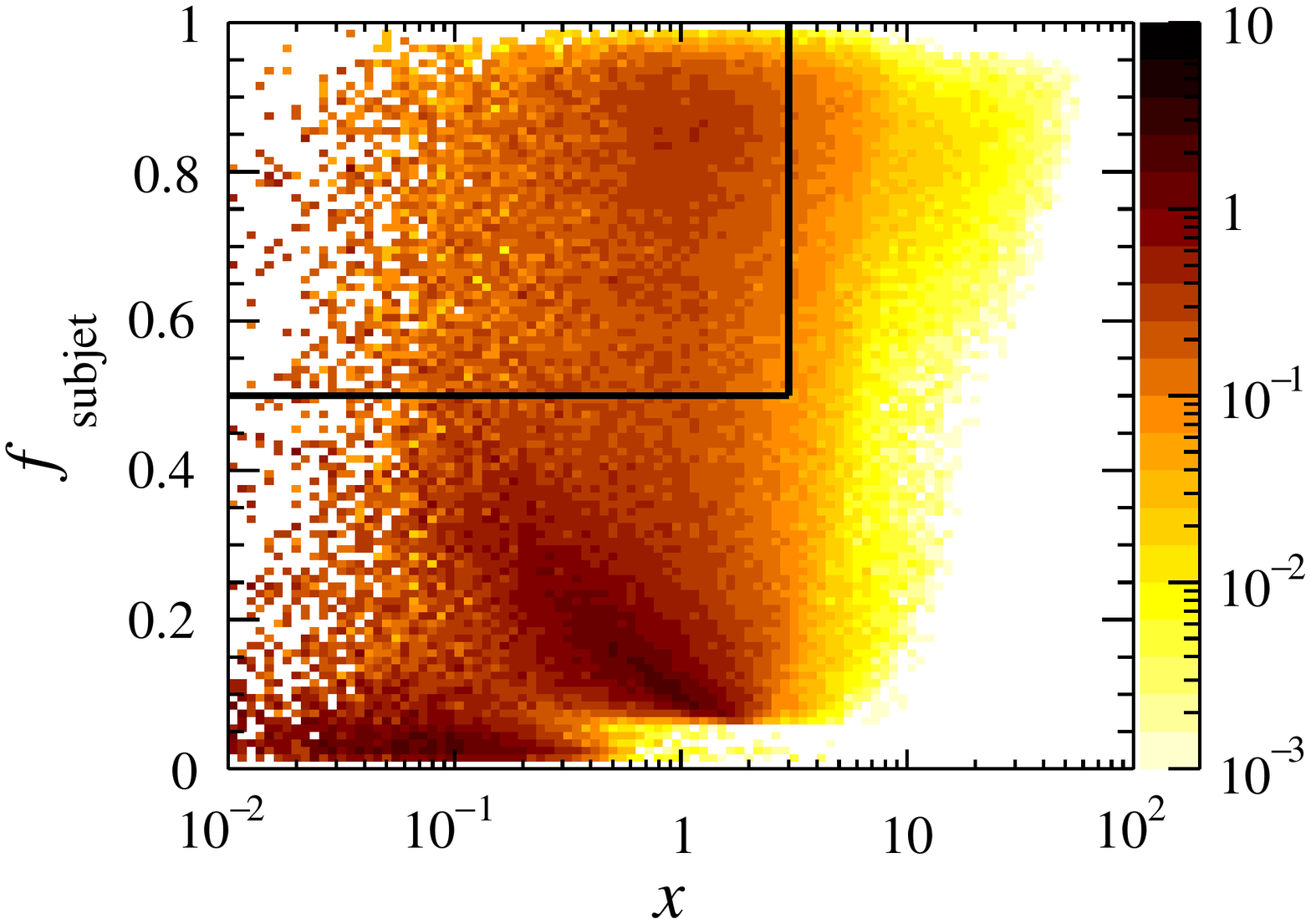}}
\caption{Density of reconstructed candidate tags with $\mu=40$ pileup
  events as a function of $\fSubjet$ vs. $x$ (summing over all $\pT$
  and $\etaJet$ bins) for (a) bottom, (b) charm, (c) light-light, and
  (d) light-heavy jets.\label{fig:xVersusF}}
\end{figure*}

The underlying physics of {\new} tagging is evident in
Fig.~\ref{fig:xVersusF}.  The $x$ distribution for \emph{bottom} jets
peaks at $x\approx0.8$ (versus $dN/dx$, which peaks around
$x\approx0.6$). This is due to a convolution of direct-$b$ and
secondary-$c$ decay, since $c$ hadron decays peak around $x=1$
(Fig.~\ref{fig:charmXF}).  In both heavy-jet classes, the $\fSubjet$
distributions favor subjets carrying nearly all of their jet's
momentum.

The $x$ distribution for \emph{light-light} jets with
sufficient $\fSubjet$ peaks to the right of $\xCut$, whereas muons
with taggable $x$ tend to be clustered into overly soft subjets.
Since \emph{light-heavy} jets contain heavy hadrons, their
high-$\fSubjet$ muons should (and do) have $b$-like values of $x$.
However, since the initial jet momentum must be shared between a pair
of heavy hadrons, many light-heavy muons with taggable $x$ fail
$\fCut$, which suppresses this background.

\section{Leptophobic $\Zp$ \label{sec:Zp}}

A simple extension of the standard model involves the addition of a
broken $\Up$ symmetry mediated by a heavy neutral $\Zp$ boson.  If the
new symmetry is associated with baryon number $B$, one would not
expect to see a dilepton signal, since only SM quarks would be
charged. To cancel anomalies, this $\Up_{B}$ should couple to
vector-like quarks, and come with at least one scalar field whose
vacuum expectation value breaks the symmetry
\cite{Dobrescu:leptophobic-Z', Dobrescu:leptophobic-anamalons}.
Assuming the vector-like quarks are kinematically inaccessible at the
LHC, a flavor-independent $\ZpB$ gauge coupling to SM quarks
\cite{Dobrescu:leptophobic-Z'}
\begin{equation}
\lagrange=\frac{\gB}{6}Z_{B\mu}^{\prime}\bar{q}\gamma^{\mu}q
\end{equation}
might lead to dijets being the only detectable signature at the LHC of
this new physics.

We would expect the purity of a dijet $\Zp$ signal to be very low,
since QCD production of dijets has an enormous cross section. This is
where {\new} tagging is useful, as the rejection of light-jet fakes
seen in Sec.\ \ref{subsec:eff} is ${\cal O}(10^3)$.  To validate our
new boosted-$b$ tag, we simulate a search for a narrow $\ZpB$ peak
above the dijet background at {\runII} of the LHC (i.e., looking for
an excess in the $d\sigma/dm_{jj}$).  We examine the experimental
reach in two dijet samples: 2-tag and 1-tag inclusive (where $N$-tag
requires at least $N$ of the top two $\pT$-ranked jets to be
{\new}-tagged).

We model $\ZpB$ production for a variety of $M_{\ZpB}$ spanning 1--4
TeV, using $pp\to\ZpB\to \pair{b}/\pair{c}(+j)$.  The optional
light-jet radiation slightly enhances the overall $\ZpB$
cross-section, but is mostly useful to improve the differential jet
distribution via MLM jet matching~\cite{Mangano:MLM-matching} in both
{\mg} and {\pythia} (in ``shower-kt'' mode~\cite{Alwall:shower-kt},
using a matching scale of $M_{\ZpB}$/20).  As before, all reconstruction
is performed using our modified {\delphes} code.

The relevant background is pure QCD, as no other SM processes contain
competing cross-sections. Both 2-tag and 1-tag backgrounds contain
$pp\to \pair{b}/\pair{c}/\pair{j}(+j)$. The 1-tag background also
includes a large contribution from $jq_{h}\to jq_{h}(+j)$ (a heavy
quark scattering off a light parton). To obtain good tagging
statistics, multiple background sets are generated, using identical
matching parameters as their corresponding signal set.

The minuscule light-jet tagging efficiency forces us to estimate the
\textit{second} tag for the 2-tag light-dijet background sample by
using our fit to the light-jet efficiency from Sec.\ \ref{subsec:eff}
as a function of jet $\pT$ and $\etaJet$. When exactly one leading jet
is tagged, we estimate the probability~$\epsilon_l$ to tag the other
jet, then re-weight the 2-tag events by a factor of
$\frac{\epsilon_l}{2(1-\epsilon_l)}$. When both leading jets are
tagged, the event is discarded, otherwise it would be double counted
by this method.

Additional cuts for our analysis include a requirement that the
pseudorapidity between jets is small $|\Delta\eta_{jj}|\le\deltaEtaMax$
in order to suppress much of the $t$-channel dijet background.  We
also require $\abs{\etaJet}\le2.7$ to ensure that both jets fall
within the muon spectrometer. While we considered including the effects
of higher order final state radiation in our mass reconstruction, we
find that adding a hard third jet to the dijet system causes an
unacceptable hardening of the QCD continuum.  Not including this
radiation, combined with the estimation of hard neutrino momenta
inherent to {\new} tagging, decreases the mass resolution of the
intrinsically narrow $\ZpB$ bosons of this model ($\Gamma_{Z^\prime} =
\frac{1}{6}\alpha_B (1+\alpha_s/\pi) M_{Z^\prime}$).  Hence, we
require a rather wide mass window $([0.85,1.25]\times M_{\ZpB})$ to
capture most of the signal.

\begin{figure*}[htb]
\subfloat[]{\includegraphics[width=0.5\textwidth]{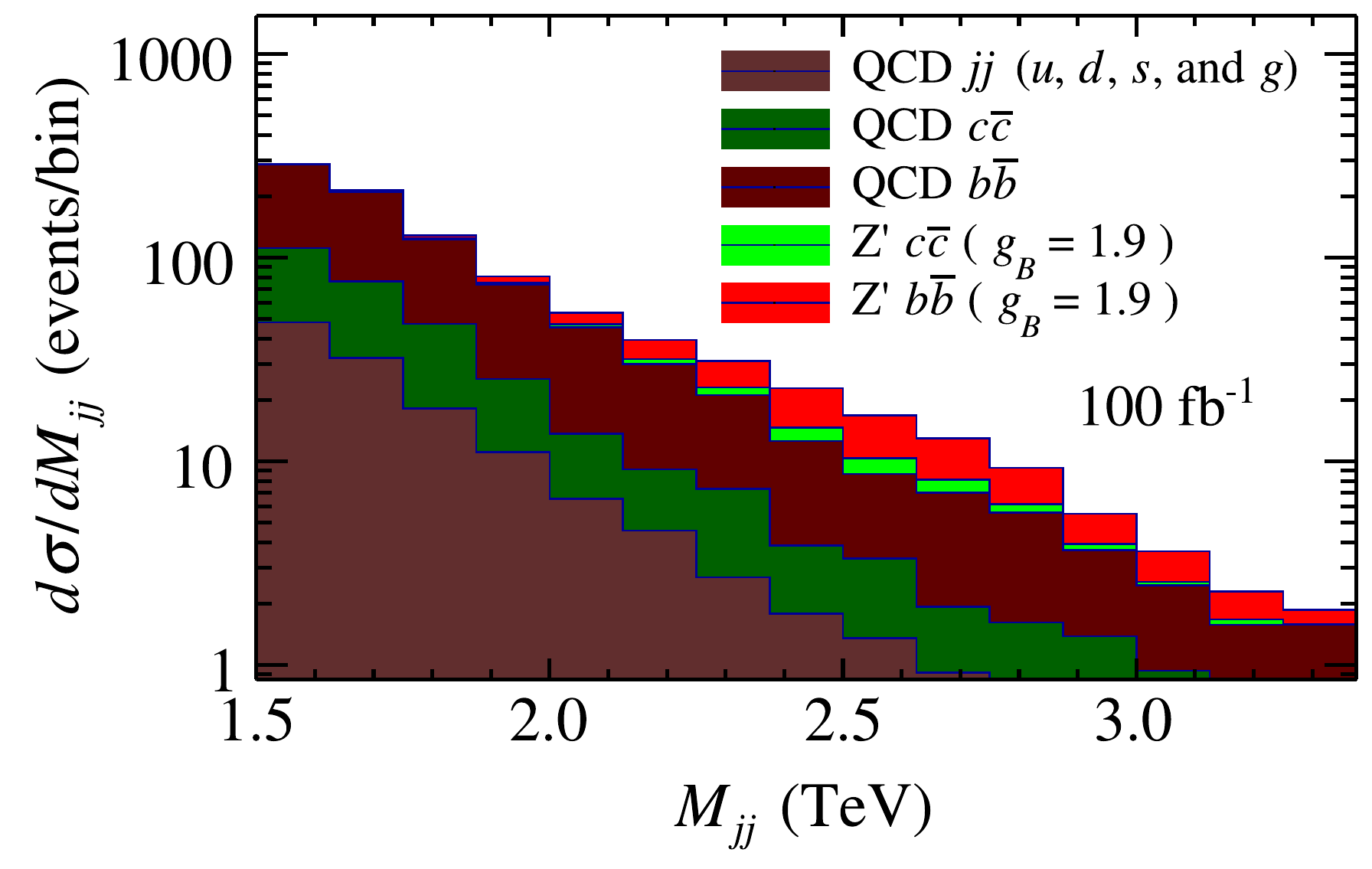}}
\subfloat[]{\includegraphics[width=0.5\textwidth]{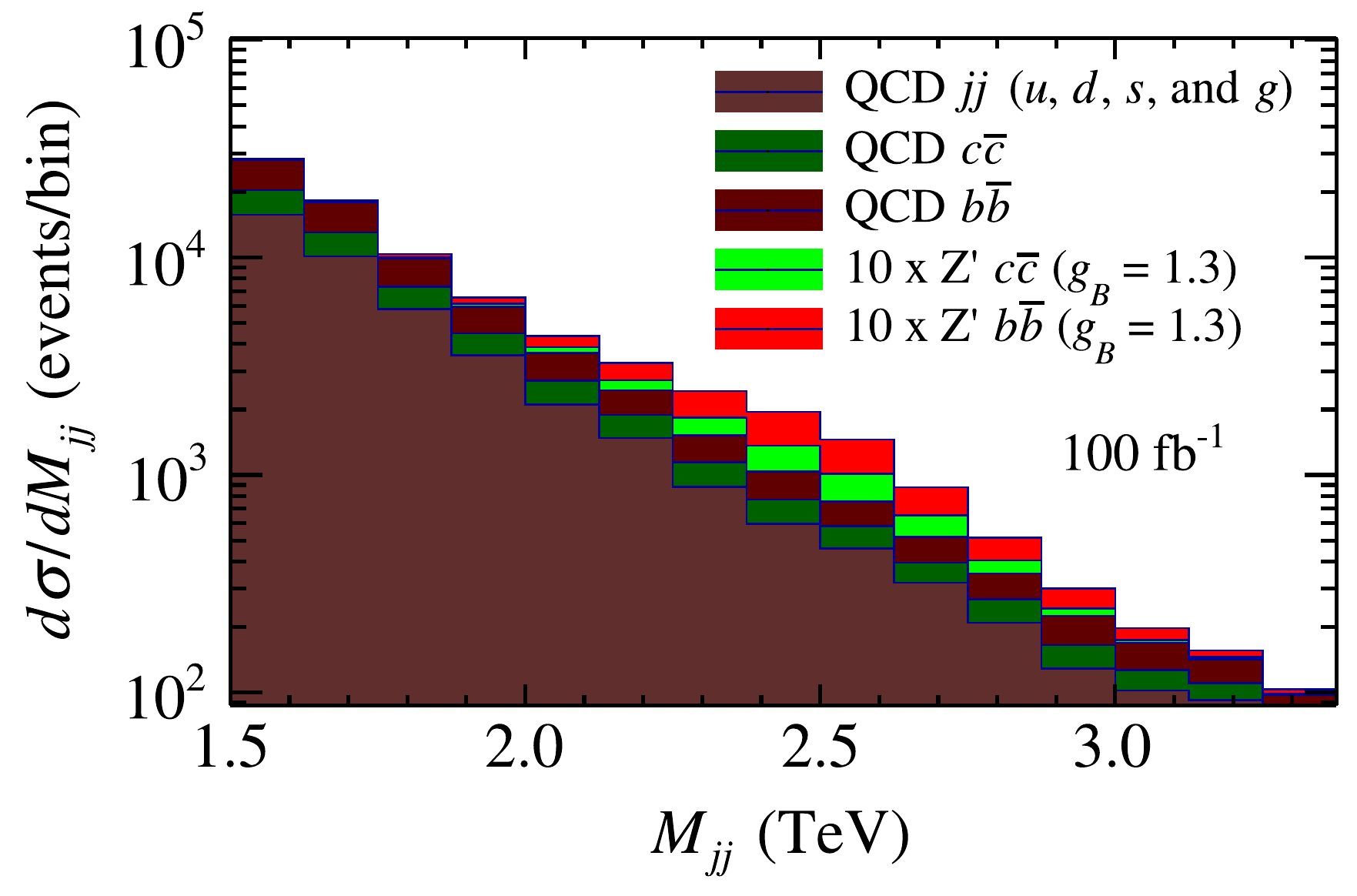}}
\caption{Events per bin expected for $5\sigma$ discovery of a 
$M_{Z^\prime_B}=2.5$ TeV signal, and backgrounds, in the (a) 2-tag and
(b) 1-tag analyses using $\runIIlum$ of integrated luminosity at
{\runII} of the LHC.
\label{fig:discovery}}
\end{figure*}

The signal and backgrounds for a $5\sigma$ discovery of a
$M_{Z^\prime_B}=2.5$ TeV $\ZpB$ boson, using our cuts for the 2-tag
and 1-tag analyses, can be seen in Fig.\ \ref{fig:discovery} for
$\runIIlum$ of integrated luminosity at the 13~TeV LHC.  The signal to
background ratio $S/B = 1/2$ for the 2-tag sample, indicating an
excellent purity.  The 1-tag sample has $S/B = 1/12$, still acceptable
given the factor of 12 more signal events that would appear in the
sample.  The peak in the 1-tag sample is slightly narrower than the
2-tag sample because only one neutrino is estimated in the boosted jet
decay.

\begin{figure}[htb]
\includegraphics[width=0.5\columnwidth]{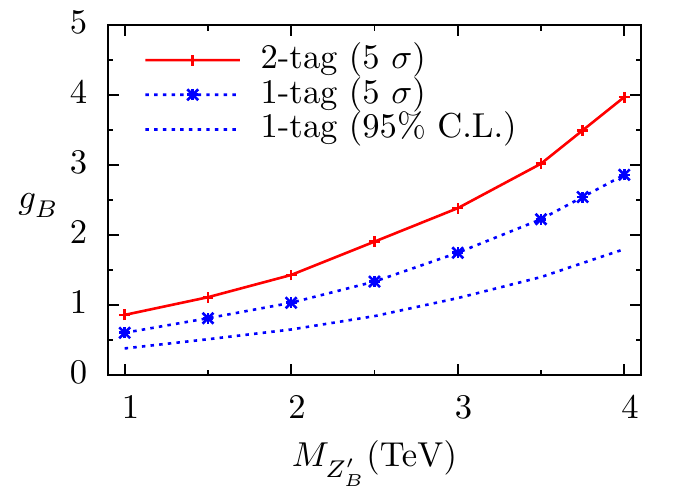}
\caption{Estimated {\runII} ($\runIIlum$) $5\sigma$ discovery potential
and $95\%$ confidence level exclusion limits for $\gB$ vs.\
$M_{Z^\prime_B}$ for the 2-tag and 1-tag analyses. 2-tag $95\%$ C.L.\ exclusion
reach (not shown) is comparable to the 1-tag discovery reach.
\label{fig:exclusions}}
\end{figure}

In Fig.~\ref{fig:exclusions} we depict the estimated discovery
potential for the 2- and 1-tag analyses, along with the 1-tag 95\%
confidence level (C.L.)  exclusion limits, for the LHC {\runII} with
the scheduled luminosity of $\runIIlum$.  Comparing our results to the
existing exclusions limits presented in Fig.\ 1 of
Ref.~\cite{Dobrescu:leptophobic-Z'}, our 2-tag discovery reach is
about 500 GeV higher in mass for large $\gB$, and is right at the
limit for smaller $\gB$.  Not shown in the Figure is the 95\% C.L.\
exclusion limit for the 2-tag search, which is slightly better than
the $5\sigma$ discovery reach in the 1-tag search.  The 1-tag search
dramatically improves the mass reach by $\sim 1.5$~TeV beyond the
current limits at large $\gB$ and, more importantly, can attain $\gB <
1$ below 2(3)~TeV for discovery(exclusion).  Hence, the {\new}
boosted-bottom tag opens a new window into leptophobic $\Zp$ boson
physics.

\section{Conclusions \label{sec:fin}}

In this paper we derive and examine the efficacy of the new {\new}
boosted-bottom-jet tag.  The {\new} tag enables a new class of high
purity searches for final states containing $b$ jets in the decays of
TeV-scale particles.  In Sec.\ \ref{sec:Zp} we propose the use of the
{\new} boosted-bottom jet tag to discover a leptophobic $\Zp$ boson at
the 13~TeV {\runII} of the LHC.  We perform two analyses based on the
number of {\new} boosted-$b$ tags: a high purity analysis with two $b$
tags, and an extended-reach analysis based on a one-tag inclusive
sample.  We find the potential for discovery of a $\ZpB$ boson with
universal coupling to quarks can be extended by 500--1000 GeV beyond
existing 95\% C.L.\ exclusion limits \cite{Dobrescu:leptophobic-Z'},
or to a factor of two smaller coupling $\gB$.

The {\new} boosted-bottom-jet tag improves on the idea of using the
angle between a muon from $B$ hadron decay and the jet centroid
proposed in Ref.\ \cite{Duffty:boostedBtag}.  By measuring the angle
between the muon and the subjet inside the $b$-jet that is likely to
contain the $B$ hadron remnant, this ``smart'' angular cut (loose for
hard muons, tight for soft muons, and scaling with the boost) obtains
efficiencies to tag $b$-jets, $c$-jets, light-jets, and light-heavy
jets (in which $g\to b\bar b$ produces real $B$ hadrons) of
$\epsilon_b = 14\%$, $\epsilon_c = 6.5\%$,
$\epsilon_{\mathrm{light-light}} = 0.14\%$, and
$\epsilon_{\mathrm{light-heavy}} = 0.5\%$, respectively. Since the
light-light fake rate is a factor of 10 smaller than current $b$ tag
estimates \cite{ATLAS:bTag-runII}, the {\new} boosted-$b$ tag should
greatly improve the uncertainties in the search for $\Wp \to t\bar b$
at ATLAS \cite{ATLAS:W'-search-8TeV, Aad:2014xra}.

Several other applications exist for {\new} boosted-$b$ tagging, such
as the search for heavy Higgs bosons in general two Higgs doublet
models with moderate $\tan(\beta)$ \cite{Craig:2015jba,Hajer:2015gka}.
When boosted-bottom tagging is combined with boosted-top tagging,
decays of heavy Higgs bosons should be accessible in the channels
$pp\to \bar{b}tH^{-}\to \bar{b}t\bar{t}b$ and
$pp\to\pair{b}H/A\to\pair{b}\pair{t}$.  Another important application
of {\new} tagging is the use of the pseudorapidity-dependent fake rate
from gluon splitting to provide an experimental handle to calibrate
this contribution to jet showering at TeV energies.

While we derive the {\new} boosted-bottom-jet tag from basic
kinematics in Sec.\ \ref{sec:tagging}, in this paper we examine its
effectiveness at the LHC in the context of the ATLAS detector.  This
choice is driven by the public ATLAS standalone/non-isolated muon
reconstruction capabilities as a function of $\pT$ and
$\eta$~\cite{ATLAS:public-muon-plots}.  We ensure that our $b$ tag is
robust in a realistic detector environment by simulating ATLAS
detector subsystems in {\delphes}, and establishing an insensitivity
to the detector details.  Given that the {\new} boosted-$b$ tag is
driven by physical principles, and not detector idiosyncrasies, we are
confident it will work as just as well with the CMS detector provided
they can reconstruct the non-isolated muons.

Naturally, the {\new} tag will require experimental validation using
heavy-flavor enriched and deficient control samples from CMS and
ATLAS.  A comparison of the {\new} tag to existing $b$ tags around
500~GeV (the lowest energy with good efficiency overlap) will permit
the extension of the {\new} tag to the highly boosted regime, where
smaller uncertainties are sorely needed.  The $b$ jet efficiency could
be extracted from $\pair{t}$ events ($\about{36\%}$ of which should
contain a semi-muonic $B$ decay).  To calibrate the light-jet fake
rate we suggest looking in a light-jet enriched dijet sample: where
one jet lacks a muon and fails a ``loose'' track-based $b$ tag, and
the other jet contains a muon.  This tag-and-probe method should
enhance the purity of the already large cross section ratio
$\sigma_{\pair{j}}/\sigma_{\pair{b}}$ by a factor of 5.

It is possible that additional improvements to the {\new} tag can be
made using capabilities specific to a given experiment.  For example,
the final layer of the ATLAS inner detector has very fine $\phi$
resolution, while the first layer of the ATLAS ECal has excellent
$\eta$ resolution.  Since the direction of the ``core'' subjet is more
important than the properties of its charged constituents (track
quantity, impact parameters, opening angles), it may be possible to
interrogate the \emph{global} nature of the core without attempting to
reconstruct its individual tracks. Given enough angular resolution, a
direct measurement of $\mCore$ could replace its manual
constraint. This procedure is essentially an extension of CMS'
particle flow algorithm to very boosted hadronic substructure.

We have described a new method for tagging boosted ($p_T > 500$ GeV)
jets that contain heavy hadrons we call {\new}-tagging.  This tag
significantly improves the purity of $b$-tagging in the boosted
regime, and will greatly extend the reach of searches for physics
beyond the standard model.  We conclude with the observation that
while we have tuned {\new}-tagging for boosted $b$ jets, the
underlying kinematics of heavy hadron decay is equally applicable to
charm jets, and a re-tuning of parameters may provide an enhanced
sensitivity to boosted-charm jet tagging.

\begin{acknowledgments}
\support
\end{acknowledgments}

\end{document}